\documentclass[english,preprint]{aastex}
\usepackage[T1]{fontenc}
\usepackage[latin9]{inputenc}
\usepackage{color}
\usepackage{amsbsy}
\usepackage{amssymb}
\usepackage{stmaryrd}
\usepackage{babel}
\begin{document}

\title{Dicke's Superradiance in Astrophysics. II \textendash{} The OH 1612
MHz Line}

\author{Fereshteh Rajabi$^{1}$ and Martin Houde$^{1,2}$}

\affil{$^{1}$Department of Physics and Astronomy, The University of Western
Ontario, London, ON, N6A 3K7, Canada}

\affil{$^{2}$Division of Physics, Mathematics and Astronomy, California
Institute of Technology, Pasadena, CA 91125, USA }
\begin{abstract}
We apply the concept of superradiance introduced by Dicke in 1954
to the OH molecule 1612 MHz spectral line often used for the detection
of masers in circumstellar envelopes of evolved stars. As the detection
of 1612 MHz OH masers in the outer shells of envelopes of these stars
implies the existence of a population inversion and a high level of
velocity coherence, and that these are two necessary requirements
for superradiance, we investigate whether superradiance can also happen
in these regions. Superradiance is characterized by high intensity,
spatially compact, burst-like features taking place over time-scales
on the order of seconds to years, depending on the size and physical
conditions present in the regions harboring such sources of radiation.
Our analysis suggests that superradiance provides a valid explanation
for previous observations of intensity flares detected in that spectral
line for the U Orionis Mira star and the IRAS18276-1431 pre-planetary
nebula. 
\end{abstract}


\section{Introduction\label{sec:Introduction}}

\begin{figure}[tb]
\epsscale{0.7}\plotone{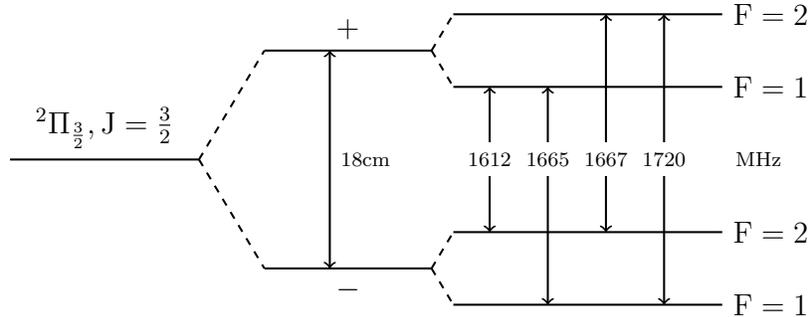}

\caption{\label{fig:OH-level diagram}The schematic diagram of the ground rotational
state of the OH molecule. The ground rotational level labelled by
$^{2}\Pi_{3/2},\:J=3/2$ splits into $\Lambda-$doublet sub-levels
shown as $\pm$ states. This splitting is due to the interaction between
the rotational and electronic angular momenta of the molecule. Each
$\Lambda-$doublet sub-level further splits into two hyperfine levels
as a result of the interaction between the electron and nuclear spins
of hydrogen atom. The four possible maser transitions are shown with
their corresponding frequencies in MHz. Note that the hyperfine splitting
is not to scale.}
\end{figure}

The OH (hydroxyl) rotational transitions at nearly 18 cm were the
first interstellar molecular lines detected in the radio range \citep{Weinreb1963,Bertolotti2015}.
The ground level of that molecule is split into two sub-levels known
as $\Lambda$-doublets with $\pm\Lambda\hbar$ energies. Each component
of the $\Lambda$-doublets is also split into two hyperfine levels
labelled $F=1$ and $F=2$, as shown in Figure \ref{fig:OH-level diagram}.
The transitions that connect sub-levels with the same $F$-values
are called the\textit{ main lines}, whereas the transitions between
sub-levels of different $F$-value are called the \emph{satellite
lines} \citep{Stahler2008}. The four transitions including the two
main lines at 1665 MHz and 1667 MHz and the two satellite lines at
1612 MHz and 1720 MHz compose the group of 18 cm wavelength lines.
In optically thin regions under conditions of local thermodynamic
equilibrium (LTE), the expected intensity ratios are approximately
1:5:9:1 for the 1612 MHz, 1665 MHz, 1667 MHz and 1720 MHz lines, respectively
\citep{Elitzur1992}. However, in several observations different line
ratios were measured \citep{McGee1965}, and in some cases the intensity
of a given line significantly exceeded that predicted using LTE \citep{Weaver1965}.
The strong anomalous line intensities were explained by postulating
maser action for the corresponding transitions.

Maser action occurs when LTE conditions are violated and velocity
coherence is achieved between a group of population-inverted molecules.
In the presence of a pumping mechanism that can maintain a higher
population in the excited level, the corresponding transition can
exhibit the exceptionally high intensity typical of maser sources
\citep{Elitzur1992}. The aforementioned study of \citet{Weaver1965}
reported the first detection of an OH maser, which was to be followed
by several other detections in different regions of the interstellar
medium (ISM). A few years later \citet{Turner1970} suggested classifying
OH maser sources into two classes: Type {\footnotesize I} and Type
{\footnotesize II}, depending on their brightest detected line. In
Type {\footnotesize I}, the main-line transitions, especially the
one at 1665 MHz, are dominant. The sources in this class are usually
detected in star-forming sites near H{\footnotesize II} regions.
Type {\footnotesize II} sources are further divided into Type {\footnotesize IIa}
and Type {\footnotesize IIb} in which the brightest line corresponds
to one of the satellite lines. Type {\footnotesize IIa} OH maser
sources, which are usually detected in supernova remnants, are brightest
at 1720 MHz, while in Type {\footnotesize IIb} the 1612 MHz line
is dominant. These sources are often spatially associated with highly
evolved stars undergoing rapid mass loss and enclosed in a circumstellar
shell \citep{Gray2012}. In addition, the first extragalactic OH maser
was detected in 1982 by \citet{Baan1982} in Arp 220 (IC 4553) with
a luminosity approximately $10^{8}$ times greater than that of typical
Galactic OH masers. This led to the term ``megamaser.'' Since then
several OH megamasers, and even gigamasers, have been detected \citep{Darling2002},
usually in the nuclear region of luminous or ultra-luminous infrared
galaxies \citep{Lockett2008}.

The existence of 18-cm OH masers confirms the possibility of inverting
populations in the ground level $\Lambda$-doublets of this molecule
in the ISM. Different pumping mechanisms are suggested for different
types of Galactic and extragalactic masers. For instance, Type {\footnotesize IIb}
maser sources near evolved stars are known to be pumped by the far-infrared
radiation emitted from dust (\citealt{Elitzur1992, He2005, Gray2005};
see Section \ref{sec:Requirements-of-Superradiance} for more details),
while it is suggested that the pumping of OH Type {\footnotesize I}
main-line masers is controlled by collisions with $\mathrm{H}_{2}$
molecules in star-forming regions\textcolor{blue}{{} }\citep{Kylafis1990}.
Maser action also requires line-of-sight velocity coherence, which
can lead to an abnormally narrow line-width through amplification
along the radiation path. For Galactic OH 18-cm lines the typical
line width is $\leq1$ km/s \citep{McBride2013}.

Population inversion and velocity coherence are also required for
the superradiance cooperative radiation process. In 1954, R. H. Dicke
pointed out that a sample consisting of $N$ excited atoms/molecules
interacting with a common radiation field cannot always be considered
as a collection of independent radiators \citep{Dicke1954}. He showed
that, under ideal conditions and after a time delay, the sample of
atoms/molecules can radiate its stored energy at an enhanced rate
$N\Gamma$; $N$ times faster than the spontaneous emission rate $\Gamma$
of a single atom/molecule. As a result, the intensity of the output
radiation $I_{\mathrm{SR}}$ scales as the square of the number of
inverted atoms/molecules $N$, unlike the linear dependency of the
radiation intensity from the corresponding non-coherent system $I_{\mathrm{nc}}$
(\citealt{Rajabi2016}; hereafter Paper I).

In the ISM, it is usually assumed that atoms/molecules interact with
the radiation field independently and the intensity of the radiation
becomes a linear function of the atomic/molecular density (if the
line is optically thin). But in the case of OH 18-cm line, for which
the detection of several maser sources verifies the possibility of
achieving population inversion and velocity coherence in some regions,
this assumption may fail and it becomes important to examine the possibility
of superradiance and coherent interactions. Accordingly, in this paper
we follow on the superradiance analysis performed in Paper I for the
H{\footnotesize I} 21 cm (magnetic dipole) transition with a similar
study for the OH 18 cm (electric dipole) line. In order to do so,
we first discuss the necessary conditions for superradiance in Section
\ref{sec:Requirements-of-Superradiance}, and narrow down our focus
to the 1612 MHz line interacting with OH molecules in the outer regions
of the circumstellar envelope (CSE) of highly evolved stars. In Section
\ref{sec:Theory}, we investigate the likelihood that these conditions
can be met in these regions using the Heisenberg approach, with a
method of analysis that is an electric dipolar version of the magnetic
dipole study found in Paper I and is similar to earlier analyses found
in the physics literature \citep{Gross1982, Benedict1996}. In Section
\ref{sec:Discussion} we discuss our numerical results on the characteristics
of a potential OH 1612 MHz coherent system, with an application to
previous observations performed on the U Orionis Mira star \citep{Jewell1981}
and the IRAS18276-1431 pre-planetary nebula \citep{Wolak2014}. We
end with a short conclusion in Section \ref{sec:Conclusion}.

\section{Requirements for Superradiance\label{sec:Requirements-of-Superradiance}}

In this section we provide a brief summary of the requirements needed
for superradiance, but a more detailed discussion will be found in
Section 2 of Paper I. Our discussion applies equally well to atoms
or molecules, however, we will focus on molecular superradiance since
OH is the subject of our analysis.

When a group of $N$ excited molecules are placed within a volume
$V$ much smaller than $\lambda^{3}$, the cube of the wavelength
of the radiation $\lambda$ interacting with the molecules, the radiation
by one molecule is seen to be in phase by the other molecules. As
a result, the radiation from the different molecules interferes constructively
and a strong directional pulse emerges from the sample \citep{Dicke1954}.
This process can also be described from the molecular point of view.
In a small-sample (i.e., defined for $V\ll\lambda^{3}$) the intermolecular
distance $r$ is much smaller than $\lambda$ and, for such small
intermolecular distances, the interaction between the molecular dipoles
and the radiation field is symmetrical throughout the sample. As a
result, after a delay time $t_{\mathrm{D}}$ the molecular dipoles
lock to a common phase and act like a single macroscopic dipole radiating
a superradiant intensity $I_{\mathrm{SR}}=NI_{\mathrm{nc}}$, where
$I_{\mathrm{nc}}$ is the intensity of a fully non-coherent system
\citep{Dicke1954, Gross1982, Benedict1996}. 

The phenomenon can be extended to a large-sample, defined as $N$
molecules distributed over a volume $V\gg\lambda^{3}$, with inter-molecular
spacings potentially larger than $\lambda$. In a large-sample, the
phase of the radiation varies from place to place as a result of propagation.
This will lead to a non-uniform spatial evolution of the molecules,
and can result in a weaker coherent behavior as compared to that of
a small-sample. However, the higher number of molecules partaking
in coherent interactions in a large-sample can make up for this, resulting
in an intense output superradiant pulse with $I_{\mathrm{SR}}=NfI_{\mathrm{nc}}$,
where the $Nf$ factor determines the enhancement of the radiation
intensity in comparison to $I_{\mathrm{nc}}$ and the efficiency of
the common phase-locking process is reflected in $f\left(<1\right)$
alone. Unlike for a small-sample, the phase-matching condition in
a large-sample can only be met in some particular directions, and
the fact that the phase-locking factor $f$ is always smaller than
unity implies a weakened superradiance. It is found that after a delay
time $\tau_{\mathrm{D}}$, a first burst of superradiance emerges
the sample followed by a number of weaker bursts, the so-called ringing
effect (see Section \ref{sec:Theory} below, and Sections 3 and 4
in Paper I). The ringing effect is associated with the re-absorption
and re-emission of radiation through the end-fire (i.e., the observer-facing
cross-section of the superradiant sample) of a large-sample interacting
with the incoming radiation originating from other parts farther away
in the sample.

Any mechanism that non-coherently reduces the excited level population
(e.g., collisional relaxation) or disturbs the phase-locking process
can diminish and even terminate superradiance. It is therefore necessary
that the time-scale of dephasing/relaxation effects be longer than
$\tau_{\mathrm{D}}$ in a large-sample (or $t_{\mathrm{D}}$ in a
small-sample) to allow for the development of any coherent behavior.
This also explains why velocity coherence is an essential condition
for superradiance. In the absence of velocity coherence, random thermal
motions in a gas results in Doppler line broadening that corresponds
to a very short dephasing time-scale (i.e., $T_{\mathrm{therm}}\ll\tau_{\mathrm{D}}$),
and renders this phenomenon the most likely cause of dephasing in
a sample. For this reason, it is not expected that superradiance could
arise in a thermally relaxed gas. Our study of superradiance is better
suited to regions of the ISM where thermal equilibrium has not been
reached (e.g., shocks; see Paper I). However, this does not imply
that Doppler broadening is not present in a velocity-coherent region,
but it is expected to be less constraining than in a thermally relaxed
environment.

For any region in the ISM, inelastic collisions with ions, electrons,
hydrogen atoms and molecules, or dust grains can further change the
internal state of an OH molecule, and if the associated time-scale
is smaller than $\tau_{\mathrm{D}}$, coherent behavior and superradiance
can be suppressed. Although they do not change the internal state
of a molecule, elastic collisions can also interrupt the coherent
phase-locking process and weaken superradiance. This is because during
an elastic collision the spacings between energy levels for the colliding
counterparts change as a result of short-range interaction forces.
After a number of such collisions each acting like a random perturbation,
the molecule can lose coherence with the interacting radiation field
\citep{Wittke1956}. Elastic collisions are normally more frequent
than inelastic collisions, and the mean time between elastic collisions
usually sets the time-scale of collisional dephasing/relaxation $T_{\mathrm{c}}$
and the corresponding condition $\tau_{\mathrm{D}}<T_{\mathrm{c}}$
is required to allow the build-up of coherent interactions.

The above discussion also implies that coherent interactions cannot
be developed in a collisionally pumped OH sample since that would
require that the pumping time-scale $T_{\mathrm{P}}=T_{\mathrm{c}}<\tau_{\mathrm{D}}$,
which contradicts the necessary condition $T_{\mathrm{c}}>\tau_{\mathrm{D}}$
for superradiance (see Paper I). Thus, for our present study of superradiance
we will only focus on OH samples that are believed to be inverted
through radiative processes. The studies of pumping mechanism of OH
masers show that the 1612 MHz masers associated with evolved stars
are pumped by far-infrared photons at 35 and 53 $\mathrm{\mu m}$
\citep{Litvak1969, Jewell1979, Gray2012, Elitzur1992}. These photons
are emitted from dust shells formed by mass losses\textbf{ }from the
central star. The radiative pumping model for 1612 MHz masers near
evolved stars is corroborated by the observation of correlated variations
in the intensities of the star and corresponding masers \citep{Harvey1974, Jewell1979}.
While recent studies by \citet{He2005} and \citet{Lockett2008} suggest
that OH megamasers, which emit primarily at 1667 MHz and 1665 MHz,
are also pumped by far-infrared radiation (more precisely 53 $\mathrm{\mu m}$
radiation from dust), the determination of the exact pumping mechanism
of main-line masers still requires more studies and collisional processes
are not generally ruled out from the pumping scenarios \citep{Gray2012}.
Hence, in this paper we limit our investigation to the possibility
of superradiance for the 1612 MHz line interacting with OH molecules
in the circumstellar envelopes of late-type stars.

\subsection{OH Samples Near Evolved Stars\label{subsec:OH-Samples-Nearby}}

One of the final stages in the evolution of a low- to intermediate-mass
star (i.e., stars with masses of about $1$ M$_{\odot}$ to $8$ M$_{\odot}$)
is the asymptotic giant branch (AGB) phase. In that stage, the star,
which is composed of an oxygen/carbon core enclosed within layers
of hydrogen and helium, becomes variable and can produce shock waves.
Shock waves initiate the mass-loss process from the photosphere of
the star to cooler regions where the gas particles can clump into
dust grains, which interact with the radiation from the star over
a broad continuum. Through these interactions the radially outgoing
photons transfer their momentum to dust grains driving them outward.
This also results in an outflow of the gas particles that are coupled
to dust grains through collisions \citep{Lamers1999,Gray2012}. As
outflowing waves move further from the star, they become cooler and
denser and form the CSE. The CSE of an evolved star can harbor masers
of different types depending on its composition. OH masers are usually
found in the CSE of oxygen-rich (or M-type) stars. Examples of such
stars are Mira variables, long-period M-type stars with periods of
100-500 days \citep{Karttunen2007}, which are known source of 1612
MHz OH maser emission. OH-IR stars are another group of long-period
variables (LPV) that were originally detected through their 1612 MHz
OH maser emission and the infrared radiation emanating from their
CSE. Miras are variable at both visible and infrared wavelengths,
whereas the CSE of OH-IR stars absorbs starlight at visible wavelengths
and re-emits it in the infrared. OH-IR stars are thought to lose mass
at a rate of $10^{-8}$ M$_{\varodot}$ yr$^{-1}$ to $10^{-4}$ M$_{\varodot}$
yr$^{-1}$ forming larger CSEs than Miras, which have a lower mass-loss
rate. The CSEs of these evolved stars are theoretically divided into
three zones \citep{Gray1999}, of which the inner and outer zones
are relevant to our study. In the outermost zone, where the CSE is
optically thin, UV light from the interstellar medium dissociates
$\mathrm{H_{2}O}$ molecules into OH. In the inner zones, the radiation
from the central star is absorbed by dust grains and is re-emitted
in the infrared. It was initially suggested by \citet{Elitzur1976}
that the 35- and 53-$\mathrm{\mu\mathrm{m}}$ infrared photons emitted
from dust pump the 1612 MHz OH masers in the outer regions of the
CSE of evolved stars. The pumping scheme proposed by \citet{Elitzur1976}
was modified by \citet{Gray2005}, and their detailed analysis of
collisional and radiative couplings of OH molecules in the expanding
CSE of OH-IR stars indicated that the strongest pumping route uses
53-$\mathrm{\mu\mathrm{m}}$ photons.

The CSE of evolved stars expands radially and reaches a constant terminal
velocity $v_{\infty}\sim10$ km $\mathrm{s}^{-1}$ to $20$ km $\mathrm{s}^{-1}$
in its outer regions \citep{Gray2012}. At distances about $r\gtrsim10^{16}$
cm from the central star, velocity coherence is achieved among OH
molecules moving with the well-defined terminal velocity in the radial
or tangential directions relative to the OH shell \citep{Draine2011}.
This velocity coherence can be maintained over the so-called Sobolev
length, which is typically on the order of $10^{14}-10^{16}$ cm for
LPV evolved stars (see Section \ref{subsec:U-Orionis} below). This,
therefore, may allow the existence of relatively long OH samples,
if inversion is achieved through infrared pumping from warm dust.
In such samples coherent correlations may develop if the shortest
non-coherent relaxation/dephasing time-scale is larger than $\tau_{\mathrm{D}}$.

Although our analysis is better adapted to regions where thermal equilibrium
has not been reached, we will approximate the time-scale of relaxation/dephasing
effects in the usual manner applicable to a thermally relaxed gas.
We should, however, keep in mind that the time-scales calculated that
way are likely to provide overestimates and should be considered as
worst case scenarios, as far as superradiance is concerned. One of
the main relaxation/dephasing mechanisms in the CSE of evolved stars
is collision. We will\textbf{ }therefore determine collision time-scales
in OH samples given a temperature $T$ and number density $n_{\mathrm{H_{2}}}$
of hydrogen molecules, which are expected to be the main collisional
partners of OH in molecular gas shells \citep{Gray2012}. These parameters
depend on the mass-loss rate of the central star in a circumstellar
envelope \citep{Goldreich1976}.\textcolor{green}{{} }For instance,
at a mass-loss rate of $1\times10^{-5}\,\mathrm{M}_{\odot}\,\mathrm{yr}^{-1}$
the abundance of OH molecules is estimated to be approximately $10$
cm$^{-3}$, while $n_{\mathrm{H_{2}}}\sim10^{6}$ cm$^{-3}$ at $r\approx10^{16}$
cm based on the study of different physical and chemical processes
taking place in the CSEs of OH-IR stars \citep{Gray2005}. Allowing
for potentially lower mass-loss rates, we will consider the 10$^{4}$
cm$^{-3}\leq n_{\mathrm{H_{2}}}\leq10^{6}$ cm$^{-3}$ range in what
follows. 

The time-scale of OH$-\mathrm{H}_{2}$ collisions can be determined
with the knowledge of the collisional cross sections in these regions.
The cross sections for inelastic OH$-\mathrm{H}_{2}$ collisions at
different temperatures are given in \citet{Offer1994}, and the related
time-scale is estimated to range from $10^{5}$ sec to $10^{7}$ sec
for 10$^{6}$ cm$^{-3}\ge n_{\mathrm{H_{2}}}\ge10^{4}$ cm$^{-3}$,
respectively. For elastic OH$-\mathrm{H}_{2}$ collisions the corresponding
time-scale is given by 

\begin{equation}
T_{\mathrm{c}}=\frac{1}{n_{\mathrm{H_{2}}}\sigma_{\mathrm{g}}\bar{v}},\label{eq:mean-time-collision}
\end{equation}
where $\sigma_{\mathrm{g}}\simeq4\times10^{-16}\,\mathrm{cm^{2}}$
is the geometrical cross-sectional area of a hydrogen molecule and
$\bar{v}$ is the mean relative velocity of OH molecules \citep{Irwin2007,Souers1986}.
For example, $\bar{v}$ is estimated to be $\sim1$ km s$^{-1}$ at
$T\sim100$ K. Inserting these values into Equation (\ref{eq:mean-time-collision})
gives $10^{4}$ sec $\leq T_{\mathrm{c}}\leq10^{6}$ sec for 10$^{6}$
cm$^{-3}\ge n_{\mathrm{H_{2}}}\ge10^{4}$ cm$^{-3}$, respectively.
As was already mentioned, elastic OH$-\mathrm{H}_{2}$ collisions
are found to be somewhat more frequent than their inelastic counterpart
and thus set the time-scale of collisional dephasing $T_{\mathrm{c}}$.

Another important process that affects the population of OH energy
levels in the CSE of evolved stars is the infrared radiative coupling
of rotational levels. As was mentioned earlier, the warm dust in the
CSE absorbs the radiation from the central star and re-emits it at
mid- and far-infrared wavelengths. The different infrared couplings
of the OH rotational levels have large transition dipoles leading
to fast excitation/relaxation rates. It should be noted that these
transition rates also depend on the opacity at the corresponding infrared
wavelengths, which vary over the different zones of the CSE. Among
the different infrared coupling routes some pump the 1612-MHz line
while others deplete the population inversion and have a relaxation
effect. The numerical studies conducted by \citet{Gray2005} show
that these infrared couplings are responsible for the inverted OH
zones, with a relaxation time-scale for the 1612-MHz line, including
collisional and radiative couplings, on the order of $\sim10^{4}$
sec for a mass loss rate of $10^{-5}$ M$_{\varodot}$ yr$^{-1}$.
Allowing for variations with different mass loss rates, we find that
the radiative relaxation time-scales are comparable to those expected
from collisions in these regions. For the sake of our discussion we
will assume, for simplicity, that the time-scale of OH$-\mathrm{H}_{2}$
collisions sets the upper limit for the characteristic time-scales
of superradiance in circumstellar OH samples, but we keep in mind
that infrared radiative coupling may also be responsible for this.
For the aforementioned range of $\mathrm{H}_{2}$ densities, when
assuming thermal equilibrium the delay time $\tau_{\mathrm{D}}$ and
the characteristic time of superradiance $T_{\mathrm{R}}$ should
not exceed $10^{4}$ sec to $10^{6}$ sec. Although these figures
should be viewed as worst case scenarios (i.e., lower limits), they
allow us to get a sense of the time-scales involved. As we will see
later, these time-scales imply intensity variations that could last
as long as several years.

\section{Analytical Model \label{sec:Theory}}

In this section we use the formalism developed in \citet{Gross1982}
to describe the behavior of a superradiant system. Since, as will
be seen in Section \ref{sec:Discussion}, the realization of superradiance
small-sample is unlikely to take place in the CSEs of evolved stars
for the collisional time-scales previously calculated (see Paper I),
we focus our analysis on the case of a large-sample.

Taking into account the dephasing/relaxation effects, the behavior
of a superradiant system can be expressed by a set of so-called Maxwell-Bloch
equations within the framework of the slowly varying envelope approximation
(SVEA) 

\begin{eqnarray}
\frac{\partial\hat{\mathbb{N}}}{\partial\tau} & = & \frac{i}{\hbar}\left(\hat{P}_{0}^{+}\hat{E}_{0}^{+}-\hat{E}_{0}^{-}\hat{P}_{0}^{-}\right)-\frac{\hat{\mathbb{N}}}{T_{1}}\label{eq:N-tau-dephase}\\
\frac{\partial\hat{P}_{0}^{+}}{\partial\tau} & = & \frac{2id^{2}}{\hbar}\hat{E}_{0}^{-}\hat{\mathbb{N}}-\frac{\hat{P}_{0}^{+}}{T_{2}}\label{eq:P+-tau-dephase}\\
\frac{\partial\hat{E}_{0}^{+}}{\partial z} & = & \frac{i\omega}{2\epsilon_{0}c}\hat{P}_{0}^{-},\label{eq:E-z-dephase}
\end{eqnarray}
where $T_{1}$ is the (phenomenological) time-scale of non-coherent
population relaxation (e.g., through inelastic collisions) and similarly
$T_{2}$ for phase relaxation (e.g., through elastic collisions).
These equations were derived within the context of the Heisenberg
representation. The quantities $\hat{P}_{0}^{\pm}$ and $\hat{E}_{0}^{\pm}$
are the envelopes for the polarization $\hat{\mathbf{P}}^{\pm}$ and
the electric field $\hat{\mathbf{E}}^{\pm}$ vectors, respectively,
which are assumed to have the following form

\begin{eqnarray}
\hat{\mathbf{P}}^{\pm}\left(z,\tau\right) & = & \hat{P}_{0}^{\pm}\left(z,\tau\right)e^{\pm i\omega\tau}\hat{\boldsymbol{\epsilon}}_{\mathrm{m}}\label{eq:P-wave}\\
\hat{\mathbf{E}}^{\pm}\left(z,\tau\right) & = & \hat{E}_{0}^{\pm}\left(z,\tau\right)e^{\mp i\omega\tau}\hat{\boldsymbol{\epsilon}}_{\mathrm{m}},\label{eq:E-wave}
\end{eqnarray}
with $\hat{\boldsymbol{\epsilon}}_{\mathrm{m}}$ the unit vector indicating
the orientation of the molecular electric dipole moment. The population
inversion density is given by (twice) $\hat{\mathbb{N}}$, while $d$
and $\omega$ are, respectively, the transition dipole matrix element
and the angular frequency of the radiation field\textbf{ }at resonance
with the molecular transition. Equations (\ref{eq:N-tau-dephase})
to (\ref{eq:E-z-dephase}) are derived using a two-level system model
and describe the evolution of the matter-field system in the retarded-time
frame $\tau$ ($=t-z/c$ , where $c$ is the speed of light). In the
ISM, an OH sample interacting with the radiation along the line-of-sight
can be modeled by a cylindrical large-sample along the $z$-axis.
Although a one-dimensional field equation discards the loss of radiation
due to diffraction or transverse effects, it is an approximation that
reduces the number of variables and allows us to move the analysis
forward while retaining the essential physics of the problem. Nonetheless,
considerations of the sample's geometry, described by the Fresnel
number $F=A/\left(\lambda L\right)$, with $A$ and $L$ the cross-section
and length of the sample, respectively, will also enter our analysis.

Under the assumption that the different dephasing time-scales are
similar (i.e., $T^{\prime}\equiv T_{1}=T_{2}$), the Maxwell-Bloch
equations can be solved by effecting the following change of variables 

\begin{eqnarray}
\hat{\mathbb{N}} & = & \frac{N}{2V}\cos\left(\theta\right)e^{-\tau/T^{\prime}}\label{eq:N_dephase}\\
\hat{P}_{0}^{+} & = & \frac{Nd}{2V}\sin\left(\theta\right)e^{-\tau/T^{\prime}},\label{eq:P_+_dephase}
\end{eqnarray}

\noindent where $\theta$ is the so-called Bloch angle and $N$ is
the number of inverted molecules at $\tau=0$ in the sample volume
$V$. Inserting Equations (\ref{eq:N_dephase}) and (\ref{eq:P_+_dephase})
into the system of Equations (\ref{eq:N-tau-dephase})-(\ref{eq:E-z-dephase})
yields

\begin{eqnarray}
\hat{E}_{0}^{+} & = & \frac{i\hbar}{2d}\frac{\partial\theta}{\partial\tau}\label{eq:E0_derivative}\\
\frac{d^{2}\theta}{dq^{2}}+\frac{1}{q}\frac{d\theta}{dq} & = & \sin\left(\theta\right),\label{eq:sine-gordon}
\end{eqnarray}
with $q$

\begin{equation}
q=2\sqrt{\frac{z\tau^{\prime}}{LT_{\mathrm{R}}}},\label{eq:newq}
\end{equation}
and $\tau^{\prime}=T^{\prime}\left(1-e^{-\tau/T^{\prime}}\right)$.
The characteristic time-scale of superradiance $T_{\mathrm{R}}$ is
given by 

\begin{equation}
T_{\mathrm{R}}=\tau_{\mathrm{sp}}\frac{8\pi}{3n\lambda^{2}L},\label{eq:TR-large-sample}
\end{equation}
where $\tau_{\mathrm{sp}}$ is the spontaneous decay time-scale of
a single molecule and $n=N/V$ the density of inverted molecules in
the sample \citep{Gross1982, Benedict1996}. Equation (\ref{eq:sine-gordon})
is the so-called Sine-Gordon equation, which can be solved numerically
to find solutions for $\theta\left(q\right)$ at the end-fire of\textbf{
}the sample (i.e., at $z=L$). The solution for $\theta$ as a function
$\tau$ can be used to evaluate $\hat{E}_{0}^{+}\left(z=L,\tau\right)$
from Equation (\ref{eq:E0_derivative}), and then the intensity of
radiation emerging from the sample with

\begin{eqnarray}
I_{\mathrm{SR}} & = & \frac{c\epsilon_{0}}{2}\left|\hat{E}_{0}^{+}\right|^{2},\label{eq:intensity}
\end{eqnarray}
where $\epsilon_{0}$ is the permittivity of vacuum. 

\begin{figure}[th]
\epsscale{0.7}\plotone{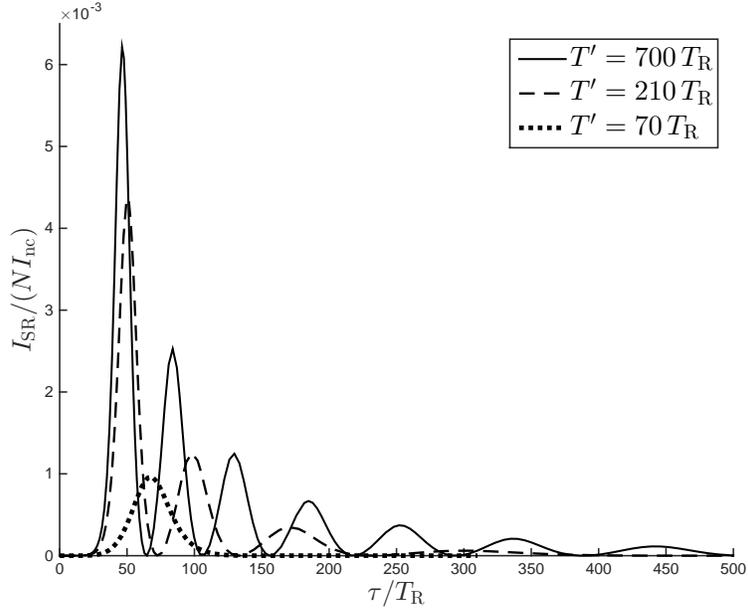}\caption{\label{fig:equaldeaphsing-largesample}The OH cylindrical large-sample.
The intensity axis, scaled to $NI_{\mathrm{nc}}$ , is plotted versus
the retarded-time axis, normalized to the superradiance characteristic
time-scale $T_{\mathrm{R}}$. The length of the sample $L$ is set
through Equation (\ref{eq:TR-large-sample}) for a given $T_{\mathrm{R}}$,
and the Fresnel number is set to unity forcing the radius of the sample
to $w=\sqrt{\lambda L/\pi}$. Dephasing effects are included for three
different time-scales $T^{\prime}=70\,T_{\mathrm{R}}$, $210\,T_{\mathrm{R}}$,
and $700\,T_{\mathrm{R}}$. }
\end{figure}

In Figure \ref{fig:equaldeaphsing-largesample} we show three solutions
for the radiation intensity from a cylindrical large-sample of OH
molecules interacting with the 1612 MHz line, for the cases where
$T^{\prime}=70\,T_{\mathrm{R}}$, $210\,T_{\mathrm{R}}$, and $700\,T_{\mathrm{R}}$.
For these calculations we set the total number density of OH molecules
to $n_{\mathrm{OH}}=10$ cm$^{-3}$, the level of inversion $\eta=0.01$,
and $T_{\mathrm{R}}=7$ days. The length of the sample $L$ is set
through Equation (\ref{eq:TR-large-sample}) for the given $T_{\mathrm{R}}$,
and the Fresnel number is set to unity forcing the radius of the sample
to $w=\sqrt{\lambda L/\pi}$. This value for the Fresnel number minimizes
diffraction losses, which are not taken into account in our model
\citep{Gross1982}. In the figure the radiation intensity axis is
scaled to $NI_{\mathrm{nc}}$ and the retarded-time axis is normalized
to $T_{\mathrm{R}}$. We can express the non-coherent intensity $I_{\mathrm{nc}}$
as

\begin{equation}
I_{\mathrm{nc}}=N\hbar\omega\left(\frac{1}{A\tau_{\mathrm{sp}}}\right)\left(\frac{\phi_{\mathrm{D}}}{4\pi}\right),\label{eq:I_nc-original-form}
\end{equation}
where $N\hbar\omega$ is the total energy initially stored in the
OH sample, of which a fraction $\phi_{\mathrm{D}}/4\pi$ emerges from
the end-fire of the sample through a cross-sectional area $A$ over
the spontaneous decay time-scale $\tau_{\mathrm{sp}}$. The superradiance
beam solid-angle $\phi_{\mathrm{D}}$ is defined as

\begin{equation}
\phi_{\mathrm{D}}=\frac{\lambda^{2}}{A},\label{eq:diffraction-angle}
\end{equation}
in the direction of the cylinder's symmetry axis where the phase-locking
condition is fulfilled. Equation (\ref{eq:I_nc-original-form}) can
be rewritten as 

\begin{equation}
I_{\mathrm{nc}}=\frac{2}{3}\frac{\hbar\omega}{AT_{\mathrm{R}}},\label{eq:I_nc_final}
\end{equation}
using Equations (\ref{eq:TR-large-sample}) and (\ref{eq:diffraction-angle}).
Equation (\ref{eq:I_nc_final}) makes it clear that the non-coherent
intensity is roughly $N$ times smaller than the maximum superradiance
intensity, since for the latter approximately $N$ inverted molecules
decay to their ground level over the characteristics time-scale of
superradiance $T_{\mathrm{R}}$.

In a large-sample, internal fluctuations (i.e., the thermal noise
or spontaneous emission) as well as an external radiation can trigger
superradiance. For the calculations presented in Figure \ref{fig:equaldeaphsing-largesample}
we used internal fluctuations to set the initial Bloch angle at $\theta_{0}=2/\sqrt{N}$
\citep{Gross1982}. Once superradiance is initiated, dipoles in a
large-sample lock into a common phase and radiate coherently after
the delay time \citep{Benedict1996}

\begin{eqnarray}
\tau_{\mathrm{D}} & \simeq & \frac{T_{\mathrm{R}}}{4}\left|\ln\left(\frac{\theta_{0}}{2\pi}\right)\right|^{2}.\label{eq:delay-time-large-sample}
\end{eqnarray}
For the above OH sample we have $\theta_{0}=4.7\times10^{-5}$ rad,
which results in $\tau_{\mathrm{D}}\simeq35\,T_{\mathrm{R}}$ using
Equation (\ref{eq:delay-time-large-sample}), in good agreement with
the time appearance of the first burst of radiation in Figure \ref{fig:equaldeaphsing-largesample}.
As can be also seen in the figure, the total energy stored in the
OH sample is released through consecutive bursts with a gradual drop
in the peak intensities, with the number of bursts depending on the
length of sample and the dephasing time-scale $T^{\prime}$. More
precisely, for longer samples and dephasing time-scales (e.g., for
$T^{\prime}=210\,T_{\mathrm{R}}$ and $700\,T_{\mathrm{R}}$ in the
figure) the process of re-absorption/re-emission takes place more
frequently at the end-fire of the sample, leading to a larger number
of burst events. However, we should also note that the (non-linear)
Sine-Gordon equation is highly sensitive to initial conditions. It
follows that the selected value for $\theta_{0}$ also has an impact
on the appearance of the intensity curve (e.g., in the number of bursts
present).

It should be also pointed out that the scaled peak intensities in
Figure \ref{fig:equaldeaphsing-largesample} indicate the phase-locking
factor $0.001\lesssim f\lesssim0.01$ depending on the dephasing time-scale.
But the large number of inverted molecules $N$ in an OH large-sample
in a circumstellar envelope will imply, even when multiplied by such
a small value for $f$, a significant enhancement factor resulting
in $I_{\mathrm{SR}}\gg I_{\mathrm{nc}}$.

\section{Discussion\label{sec:Discussion}}

The condition $\tau_{\mathrm{D}}<T^{\prime}$ further implies $T_{\mathrm{R}}<T^{\prime}$,
since in a large-sample composed of $N\gg1$ molecules $\tau_{\mathrm{D}}$
is at least an order of magnitude larger than $T_{\mathrm{R}}$ (see
Equation {[}\ref{eq:delay-time-large-sample}{]}). More precisely,
the average delay time, i.e., for several realizations of a superradiance
system with a different $\theta_{0}$, is given by \citep{Gross1982}
\begin{equation}
\left\langle \tau_{\mathrm{D}}\right\rangle =T_{\mathrm{R}}\ln\left(N\right),\label{eq:average-TD}
\end{equation}
which again indicates that $T_{R}<\left\langle \tau_{\mathrm{D}}\right\rangle $. 

In a small-sample, where there are a relatively small number of atoms,
the two time-scales $T_{\mathrm{R}}$ and $\left\langle \tau_{\mathrm{D}}\right\rangle $
are approximately of the same order of magnitude and the condition
$\tau_{\mathrm{D}}<T^{\prime}$ can be interchanged with $T_{\mathrm{R}}<T^{\prime}$.
For such cases the superradiance time-scale can be calculated from
\citep{Dicke1954}

\begin{equation}
T_{\mathrm{R}}=\frac{\tau_{\mathrm{sp}}}{\eta n_{\mathrm{OH}}V},\label{eq:TR-ideal-smallsample}
\end{equation}
where, once again, $\eta$ is the population inversion factor, $n_{\mathrm{OH}}$
is the total molecular density, and $V$ is the volume of the sample.
A small-sample of OH molecules interacting with the 1612 MHz line
is characterized by $V<\lambda^{3}\sim10^{3}$ cm$^{-3}$. Applying
this constraint on the volume and substituting $\tau_{\mathrm{sp}}\sim10^{11}$
sec for the 1612 MHz transition line transforms Equation (\ref{eq:TR-ideal-smallsample})
to 

\begin{equation}
T_{\mathrm{R}}>\frac{10^{8}}{\eta n_{\mathrm{OH}}}\,\mathrm{sec},\label{eq:TR-OH-small}
\end{equation}
yielding $T_{\mathrm{R}}>10^{9}$ sec for $n_{\mathrm{OH}}\sim10$
cm$^{-3}$ \citep{Gray2005} and $\eta\sim0.01$, which are appropriate
for masing regions in circumstellar envelopes. On the other hand,
we previously calculated the time-scale of OH$-\mathrm{H}_{2}$ collisions
$T_{\mathrm{c}}$ in circumstellar OH samples to be $\sim10^{4}$
sec to $10^{6}$ sec for molecular hydrogen densities $10^{6}$ cm$^{-3}>n_{\mathrm{H}_{2}}>10^{4}$
cm$^{-3}$ (see Section \ref{subsec:OH-Samples-Nearby}). Although
our estimated range for this collision time-scale is likely under-estimated,
it indicates that superradiance is unlikely to take place in corresponding
OH small-samples since $T_{\mathrm{c}}=T^{\prime}<T_{\mathrm{R}}$. 

In a large-sample $T_{\mathrm{R}}$ is set by the two sample parameters:
length $L$ and density of inverted molecules $n$ (see Equation {[}\ref{eq:TR-large-sample}{]}).
Although it may initially appear that in a large-sample $T_{\mathrm{R}}$
can always be set to a value smaller than $T^{\prime}$ by adjusting
$L$ or $n$, in the ISM these parameters are constrained by the physical
characteristics of the region within which the population inversion
is realized. 

In the case of CSE OH samples, the length of an inverted region depends
on the mass-loss rate of the central pulsating star, which changes
as the star evolves. The computational modeling of CSEs of OH-IR stars
by \citet{Gray2005} suggests that the radial extent of OH population-inverted
zones shrinks as the mass-loss rate of the central star increases.
More precisely, the mass-loss rate affects the optical depth of the
infrared pump photons and subsequently the thickness of inverted OH
zones. Hence, in higher mass-loss rates the envelope becomes more
opaque to pump photons and the extent of the region accessible to
pump decreases. It is expected that population-inverted regions typically
range\textcolor{white}{{} }from $10^{11}$ cm to $10^{14}$ cm\textbf{\textcolor{blue}{{}
}}in thickness \citep{Gray2005}. 

Using Equation (\ref{eq:TR-large-sample}) we can use this range for
the length $L$ of a cylindrical large-sample, along with our previous
values of $n=\eta n_{\mathrm{OH}}=0.1$ cm$^{-3}$ (i.e., 643 molecules
within $\lambda^{3}$) and $\tau_{\mathrm{sp}}=7.8\times10^{10}$
sec to find $10^{-1}\mathrm{\,sec}\gtrsim T_{\mathrm{R}}\gtrsim10^{-3}\,\mathrm{sec}$
\textcolor{black}{for}\textbf{\textcolor{black}{{} $10^{11}\,\mathrm{cm}<L<10^{14}\,\mathrm{cm}$}}\textcolor{black}{.
}Using a Fresnel number of unity to minimize diffraction losses in
our calculations, specifying a radius $w=\sqrt{\lambda L/\pi}$ ranging
from $7.7\times10^{5}$ cm to $2.4\times10^{7}$ cm, we find that
$10\,\mathrm{sec}\gtrsim\left\langle \tau_{\mathrm{D}}\right\rangle \gtrsim10^{-2}\,\mathrm{sec}$
for the same range of cylindrical lengths. The time-scales are evidently
very short in comparison to our previous estimates of $10^{4}\,\mathrm{sec}<T_{\mathrm{c}}<10^{6}\,\mathrm{sec}$.
It thus appears reasonable to expect that superradiance could take
place in the CSEs of such evolved stars. Indeed, we find that $\left\langle \tau_{\mathrm{D}}\right\rangle \sim10^{6}$
sec for as small a value as $L=10^{5}$ cm. Given these numbers, we
now investigate potential observational evidence for superradiance
in the OH 1612 MHz line.

\subsection{The U Orionis Mira Star\label{subsec:U-Orionis}}

\begin{figure}[th]
\plotone{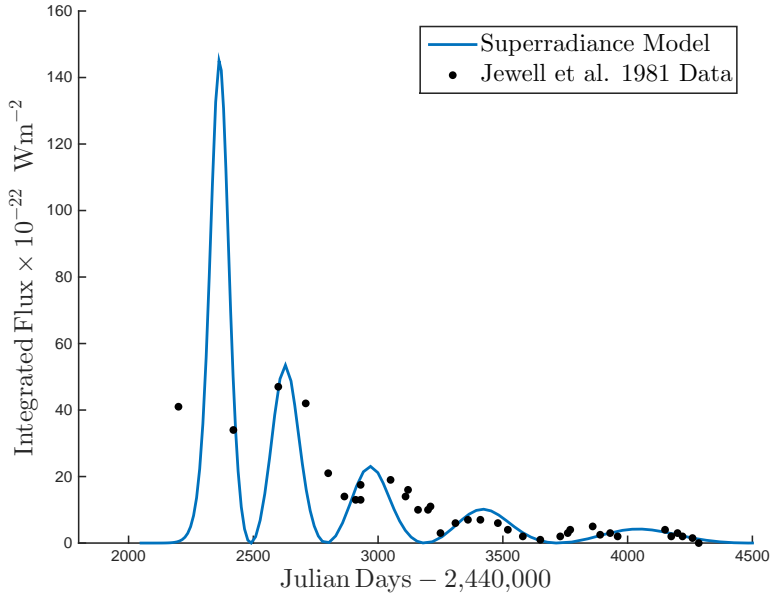}

\caption{\label{fig:Jewell}A superradiance intensity model (solid blue curve)
with $T_{\mathrm{R}}=6.5$ days, $T^{\prime}=393\,T_{\mathrm{R}}$,
and $\theta_{0}=4.4\times10^{-5}$ rad superposed on data from \citet{Jewell1981}
(black dots) obtained during the 1974 to 1979 OH 1612 MHz maser flaring
episode of U Orionis. }
\end{figure}

U Orionis is a M8 III-type OH Mira variable with a period of 372 days,
which has exhibited in the past significant variations in the intensity
of several OH maser emission lines. \citet{Jewell1981} reported the
results of their monitoring program of the 1612 MHz, 1665 MHz, and
1667 MHz masers observed in the CSE of U Ori, along with a compilation
of similar data taken from several other sources \citep{Pataki1974, Reid1977, Cimerman1979, Fix1979}
covering several years of observations (i.e., from 1974 to 1979).
All three maser lines exhibited significant flaring events during
that period. Although the 1665 MHz and 1667 MHz masers showed strong
correlation in their intensity variations, the 1612 MHz maser displayed
a completely different behavior characterized by the authors as a
``damped oscillator decline'' spanning about an order of magnitude
in intensity range over the period. These variations were also clearly
uncorrelated with the star's light curve (see Figure 1 of \citealt{Jewell1981}),
and there has been no satisfactory explanation for such behavior that
we are aware of so far.

Since the ringing in superradiance intensity displayed in the $T^{\prime}=210\,T_{\mathrm{R}}$
and $700\,T_{\mathrm{R}}$ curves shown in Figure \ref{fig:equaldeaphsing-largesample}
is also reminiscent of a ``damped oscillator'' behavior, the data
of \citet{Jewell1981} provide us with a first opportunity to test
our OH 1612 MHz superradiance model. We accordingly show in Figure
\ref{fig:Jewell} the results of our attempt.  In the figure, the
black dots are taken from Figure 1 of \citet{Jewell1981}, while the
(blue) solid curve is calculated from our model discussed in Section
\ref{sec:Theory}. We have once again used the internal fluctuation
condition to trigger superradiance (i.e, $\theta_{0}=2/\sqrt{N}$),
which resulted in $\theta_{0}=4.4\times10^{-5}$ rad ($\tau_{\mathrm{D}}\simeq35\,T_{\mathrm{R}}$;
see Equation {[}\ref{eq:delay-time-large-sample}{]}) for the chosen
parameters $T_{\mathrm{R}}=6.5$ days and $T^{\prime}=393\,T_{\mathrm{R}}$,
while keeping $n=0.1$ cm$^{-3}$. Given the simplicity of our one-dimensional
superradiance model we have not attempted to perform any formal fit
to the data, but merely adjusted the model's free parameters (i.e.,
$T_{\mathrm{R}}$ and $T^{\prime}$) to reproduce the main features
found in the data. Our model was also ``normalized'' in intensity
to that of the data. Unfortunately, the data is sparse early on (i.e.,
up to approximately Day 2600) and we cannot be certain of the proper
behavior during that period, but the intensity is well constrained
for the rest of the observation period. We should also note that the
data compiled in \citet{Jewell1981}, realized with different facilities
and instruments probing sometimes different polarization states, do
not focus on a single spectral feature but rather represent the integrated
flux over a finite bandwidth. Despite these facts and the simplicity
of our model, the oscillatory behavior of the intensity is relatively
well captured by the superradiance model. More precisely, the intensity
of the last four maxima (at times beyond Day 2500) are reasonably
well matched by the curve, both in their relative intensities and
times of occurrence. 

We note that the superradiant system stemming from our calculations
yields a cylindrical length $L=3.4\times10^{4}$ cm, which is orders
of magnitude shorter than corresponding scales expected for\textbf{
}masers. It follows that, within the context of our model, a large
number of superradiant large-samples must be responsible for creating
a radiation intensity strong enough to be detected during the flaring
period (see Section \ref{subsec:Transition-Between-Maser} below).
Also, the dephasing time $T^{\prime}$ needed to reproduce the data
corresponds to approximately 7 years ($\sim10^{8}$ sec and $\simeq10\,\tau_{\mathrm{D}}$)
and points to conditions significantly less constraining than those
previously calculated for a thermally relaxed gas, as expected. This
implies the existence of significant velocity coherence over the length
$L$ of a superradiant sample. This may not be surprising considering
the Sobolev length for this source, which we evaluate to be $L_{\mathrm{Sobolev}}\sim10^{16}$
cm using previously published data on the velocity gradient found
in U Ori's CSE \citep{Nguyen-Q-Rieu1979}. The relative smallness
of a superradiant sample (i. e., $L/L_{\mathrm{Sobolev}}\sim10^{-12}$)
is an indicator of small frequency shifts and longer dephasing time-scales
in the superradiant samples.

\subsection{The IRAS18276-1431 Pre-planetary Nebula}

\begin{figure}[th]
\plotone{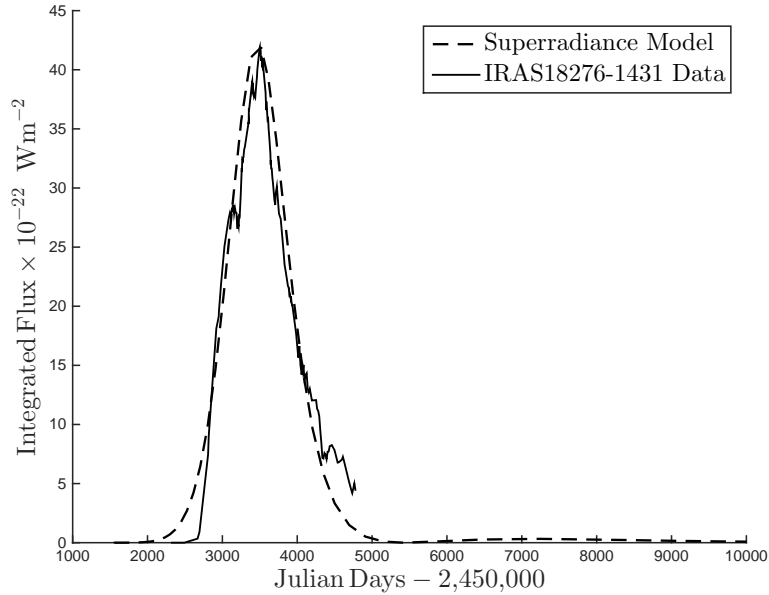}

\caption{\label{fig:Wolak}A superradiance intensity model (broken curve) with
$T_{\mathrm{R}}=42$ days, $T^{\prime}=61\,T_{\mathrm{R}}$, and $\theta_{0}=2.8\times10^{-4}$
rad superposed on data from \citet{Wolak2014} (solid curve) obtained
during the 2002 to 2009 OH 1612 MHz maser flaring episode of IRAS18276-1431. }
\end{figure}

IRAS18276-1431 (OH17.7-2.0) is a pre-planetary nebula with a detached
CSE and central star of spectral type earlier than K5, long known
for its strong 1612 MHz OH maser emission \citep{Bowers1974}. \citet{Wolak2014}
have recently published the results of a monitoring campaign performed
between 2002 and 2009, where emission from OH masers at 1612 MHz,
1665 MHz, and 1667 MHz was measured twice monthly using the Nan\c{c}ay
Radio Telescope. While a monotonic decay and rise in intensity were
detected at 1665 MHz and 1667 MHz, respectively, the integrated flux
from the red-shifted part of the 1612 MHz spectrum revealed a significant
intensity flare lasting approximately 6 years. Furthermore, during
that same period the blue-shifted part of the spectrum only displayed
a monotonic decay.

We show in Figure \ref{fig:Wolak} the results of calculations using
our superradiance model (broken curve) superposed to the data from
\citet{Wolak2014} (solid curve) for the aforementioned flaring episode
of the OH 1612 MHz line. Although the duration of the data does not
allow us to determine if the burst of radiation taking place between
approximately Day 2800 and Day 5000 is a single occurrence or part
of a series of bursts, the longer duration of the flare ($\sim2000$
days) in comparison to case of U Ori and the expected limitation on
the dephasing time $T^{\prime}$ seem to imply a single flare. Accordingly,
our superradiance model yields $T_{\mathrm{R}}=42$ days, $T^{\prime}=61\,T_{\mathrm{R}}$,
and $\theta_{0}=2/\sqrt{N}=2.8\times10^{-4}$ rad ($\tau_{\mathrm{D}}\simeq25\,T_{\mathrm{R}}$),
and provides a rewardingly nice match to the data. 

As for U Ori, we note that the superradiance cylinder length $L=5.2\times10^{3}$
cm is orders of magnitude shorter than those expected for masers,
implying that a large number of superradiant large-samples are behind
the measured intensity variation, while the required dephasing time
$T^{\prime}\simeq7$ years ($\sim10^{8}$ sec and $\simeq2.4\,\tau_{\mathrm{D}}$)
is markedly longer than the mean collision times previously calculated
for $10^{4}\:\mathrm{cm}^{-3}\le n_{\mathrm{H_{2}}}\le10^{6}$ cm$^{-3}$
at $T=100$ K, as expected. And, once again, because of the small
size of a superradiant sample ($L=5.2\times10^{3}$ cm) relative to
typical values for the Sobolev length in the CSE of evolved stars,
less constraining dephasing effects are anticipated.

\subsection{Transition Between Maser and Superradiance Modes\label{subsec:Transition-Between-Maser}}

The data sets from \citet{Jewell1981} and \citet{Wolak2014} both
indicate transitions between periods of (quasi) steady state maser
radiation and flaring episodes. Here we propose a scenario where such
transitions could take place within the context of the superradiance
model.

We know from our discussion in Section \ref{sec:Requirements-of-Superradiance}
that superradiance requires a dephasing time such that $T^{\prime}>\tau_{\mathrm{D}}$.
We therefore surmise, because of the relationship between $\tau_{\mathrm{D}}$
and $T_{\mathrm{R}}$ given in Equation (\ref{eq:delay-time-large-sample}),
the existence of a critical value $T_{\mathrm{R,crit}}$ for the characteristic
time-scale of superradiance, which cannot be exceeded for superradiance
to take place. That is, superradiance requires $T_{\mathrm{R}}\lesssim T_{\mathrm{R,crit}}$
with

\begin{equation}
T_{\mathrm{R,crit}}=\frac{4T^{\prime}}{\left|\ln\left(\frac{\theta_{0}}{2\pi}\right)\right|^{2}},\label{eq:TRcrit}
\end{equation}

\noindent which in turn can be manipulated to yield, using Equation
(\ref{eq:TR-large-sample}), a corresponding critical value for the
product of the inverted population density $n$ and the large-sample
length $L$. In other words, superradiance also implies a column density
of inverted molecules $nL\gtrsim\left(nL\right)_{\mathrm{crit}}$
with

\begin{equation}
\left(nL\right)_{\mathrm{crit}}=\frac{2\pi}{3\lambda^{2}}\frac{\tau_{\mathrm{sp}}}{T^{\prime}}\left|\ln\left(\frac{\theta_{0}}{2\pi}\right)\right|^{2}.\label{eq:nLcrit}
\end{equation}

In the AGB phase, the mass-loss rate can change significantly during
thermal pulsation periods, when the evolved star blows away its mass
in the form of super winds, or when the circumstellar envelope starts
detaching from the star. These variations happen over relatively short
time-scales (i.e., on the order of a few years) and can result in
correspondingly important changes in $L$ and $n$ in an OH large-sample.
However, variation in $nL$ do not necessarily require abrupt disruptions
in the CSE and could happen more gradually over time or even because
of local variations on smaller spatial scales. 

Whatever the case may be, we can therefore imagine a situation where
a region harboring an OH maser could experience a change in $nL$
that would push it above the critical value given in Equation (\ref{eq:nLcrit}).
At that point, the region over which $nL>\left(nL\right)_{\mathrm{crit}}$
would erupt into a superradiance mode that would overtake the maser
process, since superradiance is a lot more efficient at radiating
the energy stored in the sample (i.e., the superradiance radiation
intensity scales with $N^{2}$). The initial maser radiation field
itself could serve as a trigger for the superradiant event. Interestingly,
our superradiance models for U Ori and IRAS18276-1431 show that $nL$
(or $T_{\mathrm{R}}$ and $\tau_{\mathrm{D}}$) is only a factor of
a few (at most an order of magnitude for U Ori) higher than the critical
value.

The time-scales of the flaring events for U Ori and IRAS18276-1431
indicate, however, that the whole maser region cannot at once act
as a single coherent radiating system. Presumably some other factor
does not allow this to take place. For example, this could be because
the level of velocity coherence is not high enough over the whole
masing region to ensure that the corresponding dephasing time is sufficiently
long to allow the entire region to act as a single superradiance system.
On the other hand, velocity coherence is likely to be sufficient locally
(i.e., on scales on the order of $10^{3}$ cm to $10^{4}$ cm for
the examples considered here) to allow for the corresponding region
within the maser to break-up into a large number of smaller superradiance
large-samples. Our calculations presented in Figures \ref{fig:Jewell}
and \ref{fig:Wolak}, and their level of agreement with the corresponding
data, provide credible evidence for such a scenario.

It is interesting to note that from an observational standpoint a
single superradiant volume over which $nL>\left(nL\right)_{\mathrm{crit}}$
is unlikely to be resolvable in view of its spot size on the sky.
This would be unusual for masers, which are often resolvable through
high-resolution interferometry observations. This could also imply,
based on the model discussed here, that an unresolvable source in
a masing region exhibiting characteristics associated with superradiance
(e.g., the ringing effect in the intensity variations) could be associated
to a system composed of a single or a few superradiant sources operating
within that region. However, as is implied by the discussion above,
the converse is not necessarily true. That is, the fact that a source
is spatially resolved does not imply that it cannot consist of a group
composed of a large number of superradiant systems.

Finally, within the context of the model presented here, since a given
region can make transitions between maser and superradiance modes
we would expect that similar proper motion properties apply to both
types of sources.

\section{Conclusion\label{sec:Conclusion}}

We have applied the concept of superradiance introduced by \citet{Dicke1954}
to the OH molecule 1612 MHz spectral line often used for the detection
of masers in CSEs of evolved stars. As the detection of 1612 MHz OH
masers in the outer shells of envelopes of these stars implies the
existence of a population inversion and a high level of velocity coherence,
and that these are two necessary requirements for superradiance, we
investigated whether superradiance can also take place in these regions.
Our analysis suggests that superradiance provides a valid explanation
for previous observations of intensity flares detected in that spectral
line for the U Orionis Mira star \citep{Jewell1981} and the IRAS18276-1431
pre-planetary nebula \citep{Wolak2014}. The confirmation of superradiance
in these sources would not only reveal a new range of unexplored physical
conditions in the ISM but, on a more fundamental, also reveal the
existence of coherent quantum mechanical systems and corresponding
entangled states over length scales reaching a few times $10^{4}$
cm.

\acknowledgements{We are grateful to P. Wolak for making his data of IRAS18276-1431
available to us. M.H.'s research is funded through the NSERC Discovery
Grant and the Canada Research Chair programs.}

\end{document}